\documentstyle[epsfig,rotating]{aipproc}

\begin{document}
\title{New Particles and Interactions \\ at High Energy Muon Colliders}

\author{Stephen Godfrey}
\address{Ottawa-Carleton Institute for Physics\thanks{This Research is 
supported by the Natural Sciences and Engineering Research Council of 
Canada}, \\
Department of Physics, Carleton University, Ottawa Canada K1S 5B6}

\maketitle

\begin{abstract}
I give an overview of the ability of a high energy $\mu^+\mu^-$ 
collider to discover new particles and interactions.
I start with heavy fermions which will be the most straightforward 
to produce and observe.  I then discuss single leptoquark production 
which is produced via the quark content of the photon and the 
discovery potential for extra gauge bosons which will manifest 
themselves via deviations of observables from their standard model 
values.  Finally, contact interactions are studied as the 
generalization of looking for new interactions 
via deviations from the standard model.
\end{abstract}

\section*{Introduction}

Although the Standard Model (SM) of particle physics is in complete
agreement with present experimental data, it is believed to leave many
questions unanswered. This belief has resulted in numerous models 
that approximate the SM at presently accessable energies but which 
have a much richer particle spectrum above 100~GeV. Some models 
extend the SM gauge group by either embedding the 
extra gauge groups in a Grand Unified Group (GUT) or not embedding 
them.  GUT theories also come in supersymmetric 
varieties which leads to further phenomenological consequences, in 
particular all the supersymmetric partners of the ``conventional'' 
particles and gauge bosons \cite{susy}.   
Another broad class of models are the various 
composite models where the gauge bosons are composite, the fermions 
are composite, or the Goldstone bosons that become the longitudinal 
components of the massive gauge bosons are composite (eg. 
technicolour models).  

These models lead to many types of new particles such as; extra gauge 
bosons ($Z'$'s and $W'$'s);  new fermions which come in many 
forms such as 4th generation fermions, mirror fermions, vector 
fermions, and singlets like massive neutrinos; leptoquarks, bileptons 
and diquarks;  extended Higgs sector; excited fermions which would 
signify substructure; and other truly weird particles 
that we have yet to imagine.  

To reveal what lies beyond the SM we need to elucidate and 
complete the TeV particle spectrum.  In the remainder of this 
contribution I will survey the capability of high energy $\mu^+\mu^-$ 
colliders to discover new particles and interactions. Because this is 
such a broad topic  the survey is necessarily incomplete.  A good 
source of recent results is the contributions of the New Phenomena 
working group at the 1996 Snowmass Study on High Energy Physics 
\cite{snowmass96}.

\section*{New Fermions}

New fermions \cite{rizzo} 
are generally classified by the quantum numbers of their 
chiral components.  Fourth generation fermions are massive duplicates 
of SM fermions.  In contrast the left and right handed components of 
vector fermions are in $SU(2)_L$ and $SU(2)_R$ doublets respectively 
and mirror fermions have their left handed components in $SU(2)_L$ 
singlets and their right handed components in $SU(2)_R$ doublets.  
Except for singlet neutrinos new fermions couple to the photon 
and/or weak bosons with full strength allowing for pair production 
with unambiguous cross section.  Fermion-antifermion pairs are produced
via $\mu^+ \mu^- \to F\bar{F}$ through s-channel $\gamma$ or $Z^0$ so 
the cross section goes approximately like the QED point cross section. 
Fermions can be 
pair produced in sufficient numbers for discovery up to close to the 
kinematic limit, $\sqrt{s}/2$.

New fermions with conventional quantum numbers can mix with their SM 
partners.  The mixing is severely constrained by the non-observation 
of FCNC.  Nevertheless if the mixing is not too small new fermions can 
be produced singly in association with their light partners.  This 
results in a significantly higher search limit, 
almost $\sqrt{s}$ of the collider.


\section*{Leptoquarks}

Leptoquarks are colour triplets or anti-triplets carrying both baryon 
and lepton quantum numbers and can have spin 0 or spin 1.  They appear 
in a wide variety of models such as GUT's, technicolour, and composite 
models \cite{tom}.  
Leptoquarks reveal themselves with a dramatic signal of a 
high $p_T$ lepton balanced by a jet.

In addition to being pair produced like the fermions of the 
previous sections \cite{tom} 
leptoquarks can also be produced singly via the 
quark content of a Weissacker-Williams photon radiated off an incoming 
muon \cite{leptoq}.  The cross-section for the process is 
found by convoluting the quark distribution inside the photon with the 
$q +\mu \to LQ$ cross section:
\begin{equation}
\sigma(s) = \int f_{q/\gamma}(z,M_s^2) \hat{\sigma}(\hat{s}) dz 
             =  f_{q/\gamma}(M_s^2/s,M_s^2) 
      \frac{\mbox{$2\pi^2\kappa \alpha_{em}$}}{\mbox{$s$}}
\end{equation}
where the leptoquark couplings are replaced 
by a generic Yukawa coupling $g$ which is scaled to electromagnetic 
strength $g^2/4\pi=\kappa \alpha_{em}$.
The resulting cross-section is then convoluted with the photon 
distribution to obtain the total cross section:
\begin{eqnarray}
 \sigma(\mu^+ \mu^- & \rightarrow & X S) \nonumber \\
& = & \frac{2 \pi^2 \alpha_{em}\kappa}{s} 
    \int_{M_s^2/s}^1 \frac{dx}{x} f_{\gamma/\mu}(x,\sqrt{s}/2) 
    f_{q/\gamma}(M_s^2/(x s), M_s^2) 
\end{eqnarray}

The number of expected events ($L\times \sigma$)
for various muon collider parameters are shown in Fig. 1 where we have 
taken $\kappa=1$.  
Because this is a muon collider we are considering 2nd generation 
LQ's so that we use the $s$ and $c$-quark content 
of the photon as appropriate.  Basing discovery on the 
production of 100 LQ's leads to the search limits quoted in Table I.
The OPAL \cite{opal} and DELPHI \cite{delphi} 
collaborations have used this process to obtain limits on LQ's at LEP200.

\begin{figure}[t!] 
\centerline{
\begin{turn}{90}
\epsfig{file=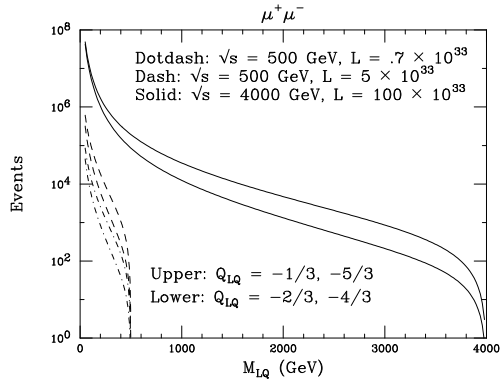,width=5.0cm,clip=}
\end{turn}
}
\vspace{10pt}
\caption[]{Event rates for leptoquark production at high energy muon 
colliders.
The results were obtained using the GRV distribution functions for the 
quark content of the photon \cite{GRV}.}
\label{fig1}
\end{figure}

\vskip 0.1cm
\begin{quote}
{\small
{\bf TABLE 1.} LQ discovery limits at $\mu^+\mu^-$ colliders for the given 
$\sqrt{s}$ and integrated luminosity.  The Scalar and Vector refers to 
the LQ spin and the $-1/3 \; -5/3$ etc. refers to its charge.
}
\end{quote}
\begin{center}
\begin{tabular}{|ll|ll|ll|}
\hline
$\sqrt{s}$ (TeV) & $L$ fb$^{-1}$ & \multicolumn{2}{|c|}{Scalar} & 
\multicolumn{2}{|c|}{Vector}  \\ \hline 
 & & -1/3, -5/3 & -4/3, -2/3 & -1/3, -5/3 & -4/3, -2/3 \\ \hline 
0.5  & 7 & 250 & 170 & 310 & 220 \\
0.5  & 50 & 400 & 310 & 440 & 360 \\
4.0 & 1000 & 3600 & 3000 & 3700 & 3400  \\ \hline
\end{tabular}
\end{center}

\vskip 0.1cm

\section*{New Gauge Bosons}

New gauge bosons are a generic prediction of models with extensions of 
the SM gauge group \cite{cvetic}.  
They contribute to $\mu^+\mu^-$ cross-sections in 
the s-channel \cite{godfrey-zp}. 
The cross-sections for various $Z'$'s are shown in Fig. 2.  It is 
clear from this figure that if production of real $Z'$'s is 
kinematically accessable it will be produced in a 
sufficiently large quantity 
so that its properties can be investigated in detail.  For the highest 
energy muon colliders being contemplated this translates into 
production of $Z'$'s of $M_{Z'}=4$~TeV (or 5~TeV depending on the 
actual $\sqrt{s}$ of the machine).  By comparison, the LHC can achieve 
a discovery reach of 4-5~TeV, depending on the specific $Z'$, based on 
roughly 10 dilepton pairs clustering at the same invariant mass.  Thus, 
the main advantage of the muon collider is that it could produce 
enough $Z'$'s to study them in detail.

\begin{figure}[t!] 
\centerline{\epsfig{file=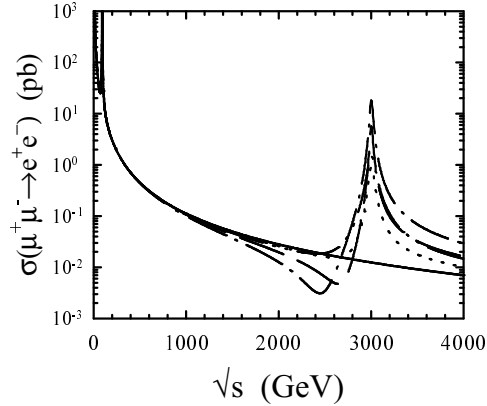,width=6.5cm,clip=}}
\vspace{10pt}
\caption[]{$\mu^+\mu^-$ cross-section as a function for $\sqrt{s}$ for 
the SM (solid), $Z_\chi$ (dashed), $Z_{LR}$ (dotted), $Z_{ALR}$ 
(dot-dashed), and $Z_{SSM}$ (dot-dot-dashed). }
\end{figure}

Searches for $Z'$s can be extended to masses much higher than 
$\sqrt{s}$ by looking for deviations from SM observables.  This is 
illustrated in the $\sigma(\mu^+\mu^- \to e^+ e^-)$ plotted in Fig. 
2 where significant deviations from the SM occur below the $Z'$ pole 
due to interference of the $Z'$ propagator with the $\gamma$ and $Z^0$ 
propagators.  

To represent a meaningful signal of new physics, deviations should be 
observed in as many observables as possible.  Observables are 
constructed from cross sections to specific final state fermions.  A 
set of such observables are; $\sigma^f$, the cross sections,  
$A_{FB}^f$, the forward-backward asymmetries, 
and $A_{LR}^f$, the left-right polarization asymmetries, 
where $f=\mu$, $\tau$, $c$, $b$, and $had=$sum over hadrons.  
To obtain discovery limits for new physics
we look for statistically significant 
deviations from standard model expectations.  In Fig. 3 a number of 
observables are shown with their standard model values and for 
various $Z'$'s as a function of the $Z'$ mass.  The 
$1-\sigma$ error bars 
shown are based on the statistics expected in the standard model. 
What is important to note is that the different observables have 
different sensitivities to different models.  For example, of the 
models shown, 
$\sigma( \mu^+ \mu^- \to e^+e^-)$ is most sensitive to $Z_{ALR}$ 
while $R^{had}$ is most sensitive to $Z_\chi$.  
Therefore to have the highest possible 
reach for the largest number of possible models
it is important to include all possible observables.
We quantify the sensitivity to an extra gauge boson by comparing the 
predictions for various observables assuming the presence of a $Z'$ to 
the predictions of the standard model 
and constructing the $\chi^2$ figure of merit. 
The ``discovery'' limits were obtained by including
the ten observables: $\sigma^\mu$, $\sigma^\tau$, 
$\sigma^c$, $\sigma^b$, $R^{had}$, $A_{FB}^\mu$, $A_{FB}^\tau$, 
$A_{FB}^c$, $A_{FB}^b$, and $P_\tau$.  In 
calculating the $\chi^2$ we assumed 35\% 
$c$-tagging efficiency and 60\% $b$-tagging efficiency.  The 99\% C.L. 
discovery limits are shown in Fig. 4 \cite{godfrey-zp}. 
Only statistical errors are 
considered in obtaining the limits shown.  We did not consider 
observables involving polarization of the initial state leptons for the 
muon colliders (although 
they were included for the $e^+e^-$ collider results).  A 
very exciting development discussed at this meeting was the 
possibility of very high muon polarization without too large a 
decrease in the luminosity.  
Polarization asymmetries are in many cases the 
most sensitive observables so that polarization is potentially very 
important for searches for $Z'$'s. 

\begin{figure}[t!]
\leavevmode
\centerline{\epsfig{file=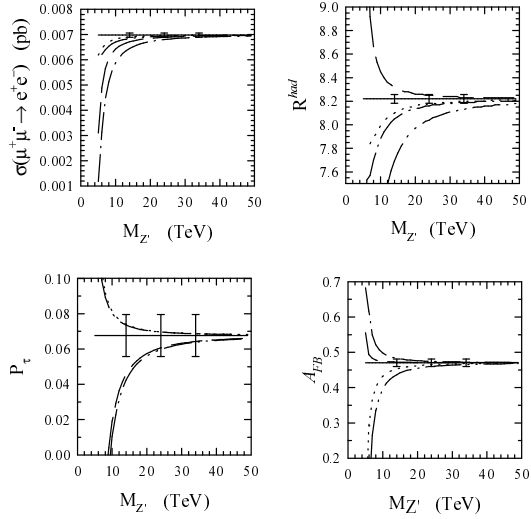,width=7.0cm,clip=}}
\caption[]{Some $\mu^+\mu^-$ observables as a function of $M_{Z'}$ 
at $\sqrt{s}=500$~GeV for the SM (solid), 
$Z_\chi$ (dashed), $Z_\eta$ (dotted), $Z_{LR}$ (dot-dashed)
and $Z_{ALR}$ (dot-dot-dashed).
The error bars are based on the statistical error assuming an integrated
luminosity of 50~fb$^{-1}$. }
\end{figure}

\begin{figure}[t!]
\leavevmode
\centerline{\epsfig{file=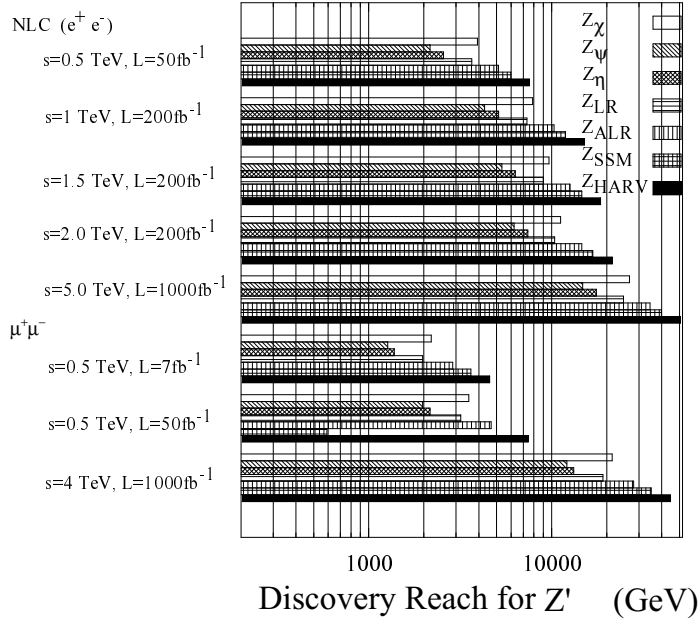,width=9.2cm,clip=}}
\caption[]{Search limits for extra neutral gauge bosons at high 
energy lepton colliders.  The criteria for obtaining these limits are 
described in the text.}
\end{figure}

\section*{Contact Interactions}

In the previous section we described how the existence of $Z'$'s might 
reveal themselves through deviations from the SM.  For very massive 
$Z'$s the $Z'$ propagator can be described by a 4-Fermi interaction
\cite{elp}:
\begin{equation}
{{g_{Z'}^2} \over {s-M_{Z'}^2}} 
\stackrel{M_{Z'}>>\sqrt{s}}{\longrightarrow} {{g_{Z'}^2} \over 
{M_{Z'}^2}} .
\end{equation}
Likewise, leptoquark exchange in the t-channel in processes like 
$\mu^+\mu^- \to q\bar{q}$ can also be described this way
\begin{equation}
{{\kappa \alpha_{em} } \over {t+M_{LQ}^2}} 
\stackrel{ M_{LQ}>>\sqrt{s}}{\longrightarrow} 
{ {\kappa \alpha_{em} } \over {M_{LQ}^2}}.
\end{equation}
Form factors or residual effective interactions 
associated with fermion substructure is also often parametrized  by 
contact terms in the low-energy Lagrangian.  Thus four fermion contact 
interactions represents a useful parametrization  of many types of new 
physics originating at a high energy scale.

These contact interactions are described by non-renormalizable operators in
the effective low-energy lagrangian.  The lowest order four-fermion contact
terms are dimension-6 and hence have dimensionful coupling constants
proportional to $g^2_{eff}/\Lambda^2$.  They are often written in the 
form \cite{elp}:
\begin{eqnarray}
{\cal L} &=&{4\pi\over 2\Lambda^2} [ \eta_{LL} (\bar{e}_L \gamma_\mu e_L)
(\bar{f}_L \gamma^\mu f_L) \nonumber \\
& + &   \eta_{LR} (\bar{e}_L \gamma_\mu e_L) (\bar{f}_R \gamma^\mu f_R)
 +  \eta_{RL} (\bar{e}_R \gamma_\mu e_R) (\bar{f}_L \gamma^\mu f_L)
\nonumber \\
& + & \eta_{RR} (\bar{e}_R \gamma_\mu e_R) (\bar{f}_R \gamma^\mu f_R) ] \,.
\end{eqnarray}  

Interference between the contact terms and the usual gauge 
interactions can lead to observable deviations from SM predictions at 
energies lower than $\Lambda$.  
The effects of a contact interaction are illustrated in Fig. 5 where
the differential cross-section for $\mu^+\mu^- \to b\bar{b}$ is 
plotted for various values of $\Lambda$.

\begin{figure}[t!]
\leavevmode
\centerline{\epsfig{file=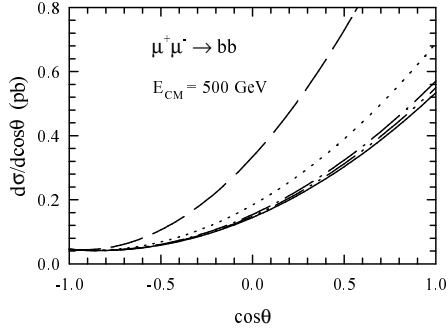,width=6.0cm,clip=}}
\caption[]{The $\cos\theta$ distribution for $\mu^+\mu^- \to b\bar{b}$ at 
$E_{CM} =0.5$ TeV  with $\eta_{LL}=+1$ for
the SM (solid), $\Lambda= 5 $~TeV (dashed), $\Lambda= 10 $~TeV (dotted),
$\Lambda= 20 $~TeV (dot-dashed), $\Lambda= 30 $~TeV (dot-dot-dashed).}
\end{figure}
 
To gauge the  sensitivity to the compositeness scale
we assume that the SM is correct and perform a $\chi^2$ analysis of the
$\cos\theta$ angular distribution.  
To perform this we choose the detector acceptance to be 
$|\cos\theta|<0.94$ (corresponding to $\theta=20^o$) \cite{cgh}.
We note that angular acceptance of a typical muon collider detector is expected
to be reduced due to additional shielding required to minimize the radiation 
backgrounds from the muon beams. 
We 
assume canonical LEP values, $\epsilon_b=25\%, \epsilon_c=5\%$ 
but warn the reader 
that these numbers are quite arbitrary and are only used for 
illustrative purposes.
We divide the angular distribution into 10 equal bins.
The $\chi^2$ distribution is evaluated by the usual expression.

The 95\% C.L. bounds on $\Lambda$ are shown graphically in Fig. 6.
Quite generally, high luminosity $\mu^+\mu^-$ colliders 
are quite sensitive to contact interactions with discovery limits 
ranging from 5 to 50 times the center of mass energy.  
As in the discussion of $Z'$'s, polarization will be important, 
especially in unravelling the chirality of deviations if they are 
observed.

\begin{figure}[h!]
\leavevmode
\centerline{\epsfig{file=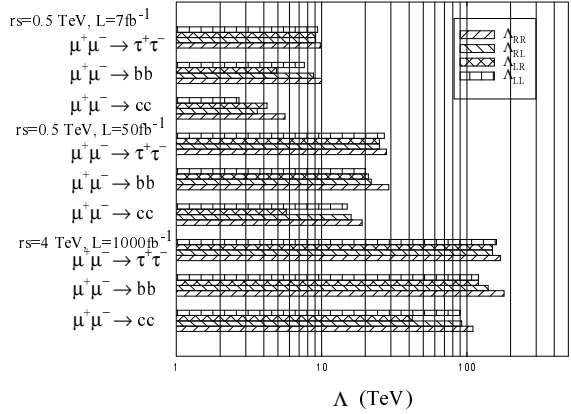,width=7.5cm,clip=}}
\caption[]{Sensitivity to the new physics scale, $\Lambda$, 
at high energy muon colliders.  The criteria for obtaining these limits are 
described in the text.}
\end{figure}

\section*{Final Comments}

The main attractions of a muon collider are its high energy reach in a 
relatively clean environment.  For certain types of physics a high 
energy muon collider could play a unique role.  For example, if the 
LHC discovered a $Z'$ with mass of 4~TeV a muon collider of 
sufficiently high energy would be able to do detailed studies of its 
properties.  Another example is the existence of heavy leptons.  These 
are notoriously difficult  and maybe impossible 
to discover at a hadron collider.  Yet for a high energy muon collider 
this would be straightforward.

The workshop discussed the likelihood of producing highely polarized 
beams.  These would play an important role in identifying the nature 
of a new particle or interaction, whether it be a leptoquark or $Z'$.  
Identification studies using polarization would be a useful excerise.  

Finally, we should keep our minds open to the possibility of genuine 
surprises which we have not yet imagined.

\end{document}